\documentclass[twoside,twocolumn,english,prb]{revtex4}
\usepackage{mathptmx}
\usepackage{helvet}
\usepackage{beramono}
\usepackage[T1]{fontenc}
\usepackage[utf8x]{inputenc}
\usepackage[letterpaper]{geometry}
\usepackage{fancyhdr}
\usepackage{babel}
\usepackage{amsmath}
\usepackage{amssymb}
\usepackage{graphicx}
\usepackage{esint}
\usepackage[unicode=true,pdfusetitle,
 bookmarks=true,bookmarksnumbered=false,bookmarksopen=false,
 breaklinks=false,pdfborder={0 0 1},backref=false,colorlinks=false]
 {hyperref}

\makeatletter

\providecommand{\tabularnewline}{\\}

\@ifundefined{textcolor}{}
{%
 \definecolor{BLACK}{gray}{0}
 \definecolor{WHITE}{gray}{1}
 \definecolor{RED}{rgb}{1,0,0}
 \definecolor{GREEN}{rgb}{0,1,0}
 \definecolor{BLUE}{rgb}{0,0,1}
 \definecolor{CYAN}{cmyk}{1,0,0,0}
 \definecolor{MAGENTA}{cmyk}{0,1,0,0}
 \definecolor{YELLOW}{cmyk}{0,0,1,0}
}

\usepackage[encapsulated]{CJK}
\usepackage{ucs}

\makeatother

\begin{document}

\title{Accurate densities of states for disordered systems from free probability:
Live Free or Diagonalize}

\author{Matthew Welborn}

\email{welborn@mit.edu}

\affiliation{Department of Chemistry, Massachusetts Institute of Technology, 77
Massachusetts Avenue, Cambridge, Massachusetts 02139, USA}

\author{Jiahao Chen \begin{CJK*}{UTF8}{bsmi}陳家豪\end{CJK*}}

\email{jiahao@mit.edu}

\affiliation{Department of Chemistry, Massachusetts Institute of Technology, 77
Massachusetts Avenue, Cambridge, Massachusetts 02139, USA}

\author{Troy Van Voorhis}

\email{tvan@mit.edu}

\affiliation{Department of Chemistry, Massachusetts Institute of Technology, 77
Massachusetts Avenue, Cambridge, Massachusetts 02139, USA}
\begin{abstract}
We investigate how free probability allows us to approximate the density
of states in tight binding models of disordered electronic systems.
Extending our previous studies of the Anderson model in one dimension
with nearest-neighbor interactions {[}J. Chen \emph{et al.}, \emph{Phys.
Rev. Lett.} \textbf{109}, 036403 (2012){]}, we find that free probability
continues to provide accurate approximations for systems with constant
interactions on two- and three-dimensional lattices or with next-nearest-neighbor
interactions, with the results being visually indistinguishable from
the numerically exact solution. For systems with disordered interactions,
we observe a small but visible degradation of the approximation. To
explain this behavior of the free approximation, we develop and apply
an asymptotic error analysis scheme to show that the approximation
is accurate to the eighth moment in the density of states for systems
with constant interactions, but is only accurate to sixth order for
systems with disordered interactions. The error analysis also allows
us to calculate asymptotic corrections to the density of states, allowing
for systematically improvable approximations as well as insight into
the sources of error without requiring a direct comparison to an exact
solution.
\end{abstract}


\maketitle

\section{Introduction}

Disordered matter is ubiquitous in nature and in manmade materials~\cite{ziman1979models}.
Random media such as glasses~\cite{Grob1976,Ford1982,Debenedetti2001},
disordered alloys~\cite{Mott1958,Tsvelick1983}, and disordered metals~\cite{Guttman1956,Dyre2000,Dugdale2005}
exhibit unusual properties resulting from the unique physics produced
by statistical fluctuations. For example, disordered materials often
exhibit unusual electronic properties, such as in the weakly bound
electrons in metal--ammonia solutions~\cite{Kraus1907,Catterall1969,Matsuishi2003},
or in water~\cite{Walker1967,Rossky1988}. Paradoxically, disorder
can also enhance transport properties of excitons in new photovoltaic
systems containing bulk heterojunction layers~\cite{Peet2009,McMahon2011,Yost2011}
and quantum dots~\cite{Barkai2004,Stefani2009}, producing anomalous
diffusion effects~\cite{Bouchaud1990,Shlesinger1993,Havlin2002}
which appear to contradict the expected effects of Anderson localization~\cite{Anderson1958,Thouless1974,Belitz1994}.
Accounting for the effects of disorder in electro-optic systems is
therefore integral for accurately modeling and engineering second--generation
photovoltaic devices~\cite{Difley2010}.

Disordered systems are challenging for conventional quantum methods,
which were developed to calculate the electronic structure of systems
with perfectly known crystal structures. Determining the electronic
properties of a disordered material thus necessitates explicit sampling
of relevant structures from thermodynamically accessible regions of
the potential energy surface, followed by quantum chemical calculations
for each sample. Furthermore, these materials lack long-range order
and must therefore be modeled with large supercells to average over
possible realizations of short-range order and to minimize finite-size
effects. These two factors conspire to amplify the cost of electronic
structure calculations on disordered materials enormously.

To avoid such expensive computations, we consider instead calculations
where the disorder is treated explicitly in the electronic Hamiltonian.
The simplest such Hamiltonian comes from the Anderson model~\cite{Anderson1958,Evers:2008p706},
which is a tight binding lattice model of the electronic structure
of a disordered electronic medium. Despite its simplicity, this model
nonetheless captures the rich physics of strong localization and can
be used to model the conductivity of disordered metals~\cite{Thouless1974,Belitz1994}.
However, the Anderson model cannot be solved exactly except in special
cases~\cite{Halperin1965,Lloyd1969}, which complicates studies of
its excitation and transport properties. Studying more complicated
systems thus requires accurate, efficiently computable approximations
for the experimental observables of interest.

Random matrix theory offers new possibilities for developing accurate
approximate solutions to disordered systems~\cite{Wigner1967,Beenakker2009,Akemann2011}.
In this Article, we focus on using random matrix theory to construct
efficient approximations for the density of states of a random medium.
The density of states is one of the most important quantities that
characterize an electronic system, and a large number of physical
observables can be calculated from it~\cite{Kittel2005}. Furthermore,
it only depends on the eigenvalues of the Hamiltonian and is thus
simpler to approximate, as information about the eigenvectors is not
needed. We have previously shown that highly accurate approximations
can be constructed using free probability theory for the simplest
possible Anderson model, i.e.\ on a one-dimensional lattice with
constant nearest-neighbor interactions~\cite{Chen2012}. However,
it remains to be seen if similar approximations are sufficient to
describe more complicated systems, and in particular if the richer
physics produced by more complicated lattices and by off-diagonal
disorder can be captured using such free probabilistic methods.

\begin{figure}[h]
\caption{\label{fig:overview}The lattices considered in this work: (a) one-dimensional
chain with nearest neighbor interactions, (b) one-dimensional chain
with many neighbors, (c) two-dimensional square lattice, (d) two-dimensional
hexagonal (honeycomb) lattice, (e) three-dimensional cubic lattice,
and (f) one dimensional-chain with disordered interactions.}
\includegraphics[width=3.5in]{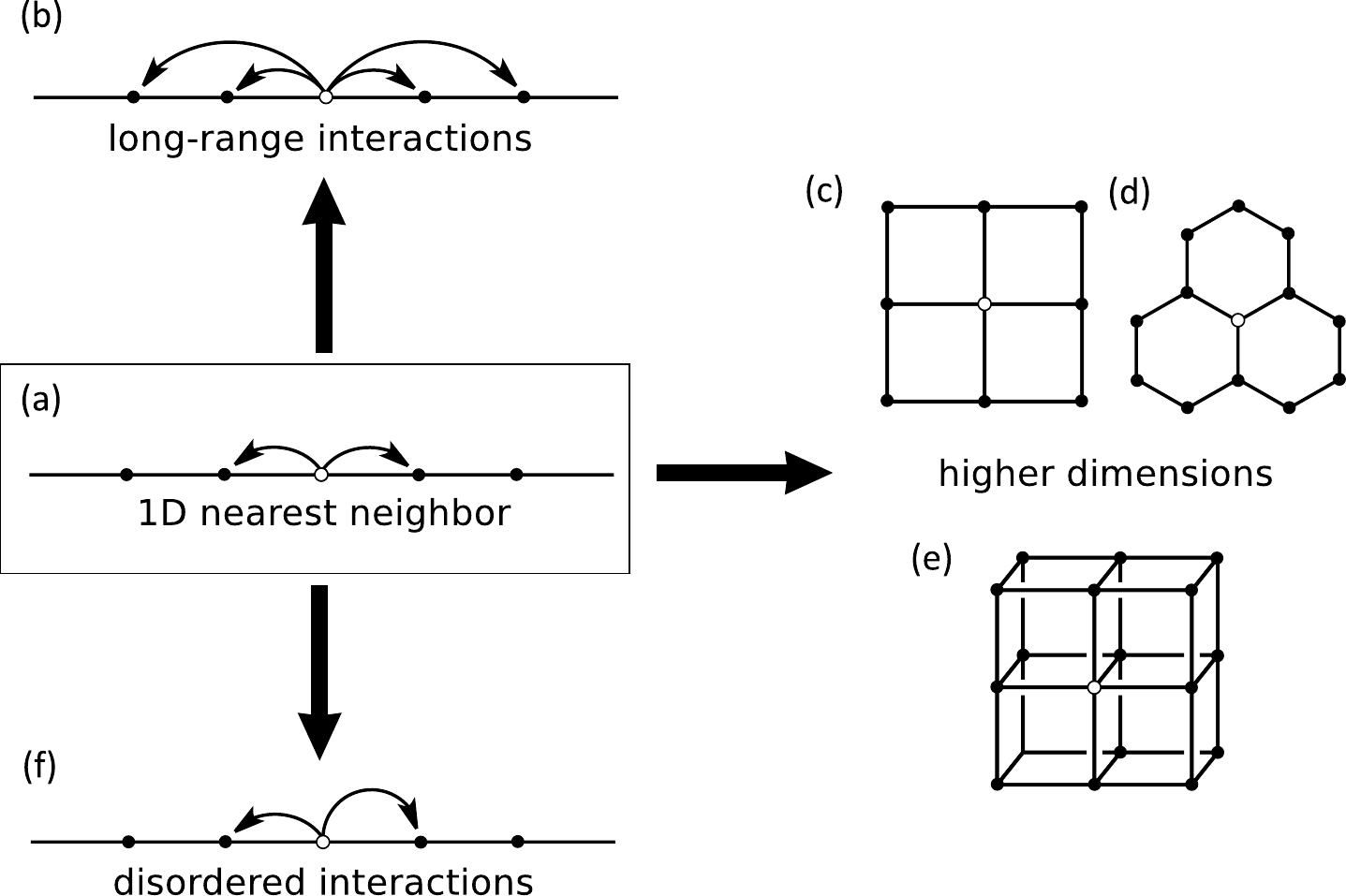}
\end{figure}

In this Article, we present a brief, self--contained introduction
to free probability theory in Section~\ref{sec:Theory}. We then
develop approximations from free probability theory in Section~\ref{sec:numbers}
that generalize our earlier study~\cite{Chen2012} in three ways.
First, we develop analogous approximations for systems with long range
interactions, specializing to the simplest such extension of a one-dimensional
lattices with next-nearest-neighbor interactions. Second, we study
lattices in two and three dimensions. We consider square and hexagonal
two-dimensional lattices to investigate the effect of coordination
on the approximations. Third, we also make the interactions random
and develop approximations for these systems as well. These cases
are summarized graphically in Figure~\ref{fig:overview} and are
representative of the diversity of disorder systems described above.
Finally, we introduce an asymptotic error analysis which allows us
to quantify and analyze the errors in the free probability approximations
in Section~\ref{sub:error}.

\section{Free probability\label{sec:Theory}}

\subsection{Free independence}

In this section, we briefly introduce free probability by highlighting
its parallels with (classical) probability theory. One of the core
ideas in probability theory~\cite{Feller1971} is how to characterize
the relationship between two (scalar-valued) random variables $x$
and $y$. They may be correlated, so that the joint moment $\left\langle xy\right\rangle $
is not simply the product of the individual expectations $\left\langle x\right\rangle \left\langle y\right\rangle $,
or they may be correlated in a higher order moment, i.e.\ there are
some smallest positive integers $m$ and $n$ for which $\left\langle x^{m}y^{n}\right\rangle \ne\left\langle x^{m}\right\rangle \left\langle y^{n}\right\rangle $.
If neither case holds, then they are said to be independent, i.e.\ that
all their joint moments of the form $\left\langle x^{m}y^{n}\right\rangle $
factorize into products of the form $\left\langle x^{m}y^{n}\right\rangle =\left\langle x^{m}\right\rangle \left\langle y^{n}\right\rangle $.
For random matrices, similar statements can be written down if the
expectation $\left\langle \cdot\right\rangle $ is interpreted as
the normalized expectation of the trace, i.e.\ $\left\langle \cdot\right\rangle =\frac{1}{N}\mathbb{E}\mbox{Tr }\cdot$,
where $N$ is the size of the matrix. However, matrices in general
do not commute, and therefore this notion of independence is no longer
unique: for noncommuting random variables, one cannot simply take
a joint moment of the form $\left\langle A^{m_{1}}B^{n_{1}}\cdots A^{m_{k}}B^{n_{k}}\right\rangle $
and assert it to be equal in general to $\left\langle A^{m_{1}+\dots+m_{k}}B^{n_{1}+\dots+n_{k}}\right\rangle $.
The complications introduced by noncommutativity give rise to a different
theory, known as free probability theory, for noncommuting random
variables~\cite{Nica2006}. This theory introduces the notion of
free independence, which is the noncommutative analogue of (classical)
independence. Specifically, two noncommutative random variables $A$
and $B$ are said to be freely independent if for all positive integers
$m_{1}$,...,$m_{k}$, $n_{1}$,...,$n_{k}$, the centered joint moment
vanishes, i.e. 
\begin{equation}
\left\langle \overline{A^{m_{1}}}\ \overline{B^{n_{1}}}\cdots\overline{A^{m_{k}}}\ \overline{B^{n_{k}}}\right\rangle =0,\label{eq:cjm}
\end{equation}
where we have introduced the centering notation $\overline{A}=A-\left\langle A\right\rangle $.
This naturally generalizes the notion of classical independence to
noncommuting variables, as the former is equivalent to requiring that
all the centered joint moments of the form $\left\langle \overline{x^{m}}\ \overline{y^{n}}\right\rangle $
vanish. If the expectation $\left\langle A\right\rangle $ is reinterpreted
as the normalized expectation of the trace of a random matrix $A$,
then the machinery of free independence can be applied directly to
random matrices~\cite{Voiculescu1991}.

\subsection{Free independence and the $R$-transform\label{sub:Free-independence}}

One of the central results of classical probability theory is that
if $x$ and $y$ are independent random variables with distributions
$p_{X}\left(x\right)$ and $p_{Y}\left(y\right)$ respectively, then
the probability distribution of their sum $x+y$ is given by the convolution
of the distributions, i.e.~\cite{Feller1971}
\begin{equation}
p_{X+Y}\left(y\right)=\int_{-\infty}^{\infty}p_{X}\left(x\right)p_{Y}\left(x-y\right)dx.
\end{equation}
An analogous result holds for freely independent noncommuting random
variables and is known as the (additive) free convolution; this is
most conveniently defined using the $R$-transform~\cite{Nica2006,Voiculescu1986,Speicher1994}.
For a probability density $p(x)$ supported on $\left[a,b\right]$,
its $R$-transform $R\left(w\right)$ is defined implicitly via\begin{subequations}

\begin{align}
G(z) & =\lim_{\epsilon\rightarrow0^{+}}\int_{a}^{b}\frac{p(x)}{z-\left(x+i\epsilon\right)}dx\\
R(w) & =G^{-1}(w)-\frac{1}{w}.
\end{align}
\end{subequations}These quantities have natural analogues in Green
function theory: $p\left(x\right)$ is the density of states, i.e.\ the
distribution of eigenvalues of the underlying random matrix; $G$$\left(z\right)$
is the Cauchy transform of $p\left(x\right)$, which is the retarded
Green function; and $G^{-1}\left(w\right)=R\left(w\right)+1/w$ is
the self-energy. The $R$-transform allows us to define the free convolution
of $A$ and $B$, denoted $A\boxplus$B, by adding the individual
$R$-transforms
\begin{equation}
R_{A\boxplus B}\left(w\right):=R_{A}\left(w\right)+R_{B}\left(w\right).
\end{equation}
This finally allows to state that if $A$ and $B$ are freely independent,
then the sum $A+B$ must satisfy
\begin{equation}
R_{A+B}\left(w\right)=R_{A\boxplus B}\left(w\right).\label{eq:free-indep-R}
\end{equation}

In general, random matrices $A$ and $B$ are neither classically
independent nor freely independent. However, we can always construct
combinations of them that are always freely independent. One such
combination is $A+Q^{\dagger}BQ$, where $Q$ is a random orthogonal
(unitary) matrix of uniform Haar measure, as applied to real symmetric
(Hermitian) $A$ and $B$~\cite{Edelman2005}. The similarity transform
effected by $Q$ randomly rotates the basis of $B$, so that the eigenvectors
of $A$ and $B$ are always in generic position, i.e.\ that any eigenvector
of $A$ is uncorrelated with any eigenvector of $B$~\cite{Akemann2011}.
This is the main result that we wish to exploit. While in general
$A$ and $B$ are not freely independent, and hence (\ref{eq:free-indep-R})
fails to hold exactly, we can nonetheless make the approximation that
(\ref{eq:free-indep-R}) holds \emph{approximately}, and use this
as a way to calculate the density of states of a random matrix $H$
using only its decomposition into a matrix sum $H=A+B$. Our application
of this idea to the Anderson model is described below.

\section{Numerical results\label{sec:numbers}}

\subsection{Computation of the Density of States and its Free Approximant}

We now wish to apply the framework of free probability theory to study
Anderson models beyond the one-dimensional nearest-neighbor model
which was the focus of our initial study~\cite{Chen2012}. It is
well-known that more complicated Anderson models exhibit rich physics
that are absent in the simplest case. First, the one-dimensional Anderson
Hamiltonian with long range interactions has delocalized eigenstates
at low energies and an asymmetric density of states, features that
are absent in the simplest Anderson model~\cite{Cressoni1998,DeBrito2004,DeMoura2005,Malyshev2004,Rodriguez2003}.
These long range interactions give rise to slowly decaying interactions
in many systems, such as spin glasses~\cite{Ford1982,Binder1986}
and ionic liquids~\cite{Pitzer1985}. Second, two-dimensional lattices
can exhibit weak localization~\cite{Abrahams1979}, which is responsible
for the unusual conductivities of low temperature metal thin films~\cite{Dolan1979,Bergmann1984}.
The hexagonal (honeycomb) lattice is of particular interest as a tight
binding model for nanostructured carbon allotropes such as carbon
nanotubes~\cite{Saito1999} and graphene~\cite{Hobson1952,CastroNeto2009},
which exhibit novel electronic phases with chirally tunable band gaps~\cite{Hamada1992,Samarakoon2010}
and topological insulation~\cite{CastroNeto2006,Moore2010}. Third,
the Anderson model in three dimensions exhibits nontrivial localization
phases that are connected by the metal--insulator transition~\cite{Anderson1958,Schonhammer1973}.
Fourth, systems with off-diagonal disorder, such as substitutional
alloys and Frenkel excitons in molecular aggregates~\cite{ziman1979models,Fidder1991},
exhibit rich physics such as localization transitions in lattices
of any dimension~\cite{Antoniou1977}, localization dependence on
lattice geometry~\cite{Hu1984}, Van Hove singularities~\cite{Brezini1990},
and asymmetries in the density of states~\cite{Fidder1991}. Despite
intense interest in the effects of off-diagonal disorder, such systems
have resisted accurate modeling~\cite{Blackman1971,Koepernik1998,Esterling1975,Elliott1974,Tanaka1971,Bishop1974,Shiba1971,Schwartz1972,Schwartz1973}.
We are therefore interested to find out if our approximations as developed
in our initial study~\cite{Chen2012} can be applied also to all
these disordered systems.

The Anderson model can be represented in the site basis by the matrix
with elements
\begin{equation}
H_{ij}=g_{i}\delta_{ij}+J_{ij}
\end{equation}
where $g_{i}$ is the energy of site $i$, $\delta_{ij}$ is the usual
Kronecker delta, and $J_{ij}$ is the matrix of interactions with
$J_{ii}=0$. Unless otherwise specified, we further specialize to
the case of constant interactions between connected neighbors, so
that $J_{ij}=JM_{ij}$ where $J$ is a scalar constant representing
the interaction strength, and $M$ is the adjacency matrix of the
underlying lattice. Unless specified otherwise, we also apply vanishing
(Dirichlet) boundary conditions, as this reduces finite-size fluctuations
in the density of states relative to periodic boundary conditions.
For concrete numerical calculations, we also choose the site energies
$g_{i}$ to be iid Gaussian random variables of variance $\sigma^{2}$
and mean 0. With these assumptions, the strength of disorder in the
system can be quantified by a single dimensionless parameter $\sigma/J$. 

The particular quantity we are interested in approximating is the
density of states, which is one of the most important descriptors
of electronic band structure in condensed matter systems~\cite{Kittel2005}.
It is defined as the distribution
\begin{equation}
\rho_{H}\left(x\right)=\left\langle \sum_{j}\delta\left(x-\epsilon_{j}\right)\right\rangle 
\end{equation}
where $\epsilon_{j}$ is the $j$th eigenvalue of a sample of $H$
and the expectation $\left\langle \cdot\right\rangle $ is the ensemble
average.

To apply the approximations from free probability theory, we partition
our Hamiltonian matrix into its diagonal and off--diagonal components
$A$ and $B$. The density of states of $A$ is simply a Gaussian
of mean 0 and variance $\sigma^{2}$, and for many of our cases studied
below, the density of states of $B$ is proportional to the adjacency
matrix of well--known graphs~\cite{Strang1999} and hence is known
analytically. We then construct the free approximant
\begin{equation}
H^{\prime}=A+Q^{T}BQ\label{eq:free-approximant}
\end{equation}
where $Q$ is a random orthogonal matrix of uniform Haar measure as
discussed in Section~\ref{sub:Free-independence}, and find its density
of states $\rho_{H^{\prime}}$. Specific samples of $Q$ can be generated
by taking the orthogonal part of the QR decomposition~\cite{Golub1996}
of matrix from the Gaussian orthogonal ensemble (GOE)~\cite{Diaconis2005}.
We then average the approximate density of states over many realizations
of the Hamiltonian and $Q$ and compare it to the ensemble averaged
density of states generated from exact diagonalization of the Hamiltonian.
We choose the number of samples to be sufficient to converge the density
of states with respect to the disorder in the Hamiltonian. While this
is not the most efficient way of computing free convolutions, it provides
a general and robust test for the quality of the free approximation.
The free approximant can be computed efficiently using numerical free
convolution techniques~\cite{Rao2012}.

\subsection{One-dimensional chain\label{sub:1dTB}}

\begin{figure*}[t]
\caption{\label{fig:alldata}Comparison of exact density of states (lines)
with free probability approximant (circles) for the lattices in Figure~\ref{fig:overview},
with (b) showing the case of $n=4$ neighbors. Data are shown for
multiple values of a dimensionless parameter quantifying the strength
of disorder to show the robustness of this approximation. For (a-e),
this parameter is $\sigma/J$, the ratio of the noisiness of diagonal
elements to the strength of off-diagonal interaction. In (f), the
axis is chosen to be the relative strength of off-diagonal disorder
to diagonal disorder, $\sigma^{*}/\sigma$, with $\sigma/J=1$.}
\includegraphics[width=1\textwidth]{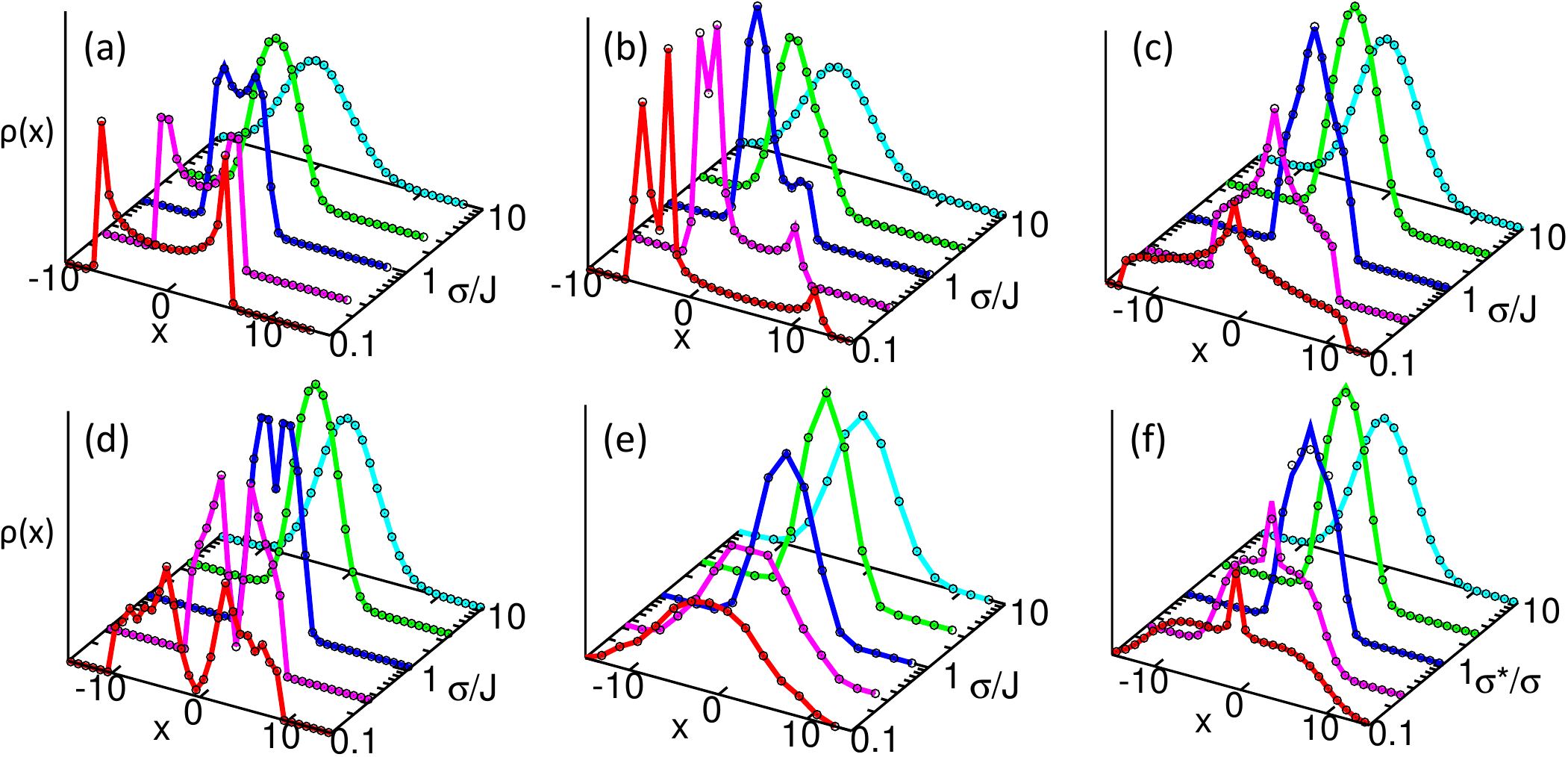}
\end{figure*}

We now proceed to apply the theory of the previous section to specific
examples of the Anderson model on various lattices. Previously, we
had studied the Anderson model on the one-dimensional chain~\cite{Chen2012}: 

\begin{equation}
H_{ij}=g_{i}\delta_{ij}+J\left(\delta_{i,j+1}+\delta_{i,j-1}\right),\label{eq:anderson}
\end{equation}
which is arguably the simplest model of a disordered system. Despite
its simple tridiagonal form, this Hamiltonian does not have an exact
solution for its density of states, and many approximations for it
have been developed~\cite{Yonezawa1973}. However, unlike the original
Hamiltonian, the diagonal and off-diagonal components each have a
known density of states when considered separately. To calculate the
density of states of the Hamiltonian, we diagonalized 1000 samples
of $1000\times1000$ matrices, so that the resulting density of states
is converged with respect to both disorder and finite-size effects.
The results are shown in Figure~\ref{fig:alldata}(a), demonstrating
that the free approximation to the density of states is visually indistinguishable
from the exact result over all the entire possible range of disorder
strength $\sigma/J$.

\subsection{One-dimensional lattice with non-neighbor interactions\label{sub:1dnnn}}

Going beyond tridiagonal Hamiltonians, we next study the Anderson
model on a one-dimensional chain with constant interactions to $n$
neighbors. The Hamiltonian then takes the form:

\begin{equation}
H_{ij}^{1D}=g_{i}\delta_{ij}+J\left[\sum_{k=1}^{n}\delta_{i,j+k}+\delta_{i+k,j}\right].\label{eq:1dnnn}
\end{equation}
where we use the superscript to distinguish the one-dimensional many-neighbor
Hamiltonian from its higher dimensional analogs. Unlike the nearest-neighbor
interaction case above, the density of states is known to exhibit
Van Hove singularities at all but the strongest disorder~\cite{Thouless1974,ziman1979models}.

We average over 1000 samples of $1000\times1000$ Hamiltonians, which
as before ensures that the density of states is numerically converged
with respect to statistical fluctuations and finite-size effects.
We looked at the case of $n=2,...,6$ neighbors with identical interaction
strengths, and also interaction strengths that decayed linearly with
distance to better model the decay of interactions with distance in
more realistic systems. The free approximant is of similar quality
in all cases. As shown in Figure~\ref{fig:alldata}(b) for $n=4$
neighbors, the free approximant reproduces these singular features
of the density of states, unlike perturbative methods which are known
to smooth them out~\cite{Allan1984,Haydock1980}. The reproduction
of singularities by the free approximant parallels similar observations
found in other applications of free probability to quantum information
theory~\cite{Movassagh2011a}.

\subsection{Square, hexagonal and cubic lattices\label{sub:2d1nn}}

We now investigate the effect of dimensionality on the accuracy of
the free approximant in three lattices. First, we consider the Anderson
model on the square lattice, with Hamiltonian:

\begin{equation}
H^{2D}=B^{1D}\otimes I+I\otimes B^{1D}+A
\end{equation}
where $B^{1D}$ is the off-diagonal part of the $H^{1D}$ defined
in equation (\ref{eq:1dnnn}), $I$ is the identity matrix with the
same dimensions as $B^{1D}$, $A$ is the diagonal matrix of independent
random site energies of appropriate dimension, and $\otimes$ is the
Kronecker (direct) product. We have found that a square lattice of
$50\times50=2500$ sites is the smallest lattice with negligible finite
size fluctuations in the density of states. As such, we calculated
the density of states for 500 samples of $3600\times3600$ Hamiltonians.
We find that for both nearest-neighbors (shown in Figure \ref{fig:alldata}(c))
and non-nearest-neighbors (specifically, $n=2,...,6$), the free approximation
is again visually identical to the exact answer.

Second, we consider the honeycomb (hexagonal) lattice, which has a
lower coordination number than the square lattice. Its adjacency matrix
does not have a simple closed form, but can nonetheless be easily
generated. For this lattice, we averaged over 1000 samples of matrices
of size $968\times968$, and applied periodic boundary conditions
to illustrate their effect. As in the square case of the two dimensional
grid model, the density of states of the honeycomb lattice with any
number of coupled neighbor shells is well reproduced by the free approximant
(Figure \ref{fig:alldata}(d)), even reproducing the Van Hove singularities
at low to moderate site disorder. Additionally, we see that the finite-size
oscillations at low disorder ($\sigma/J\sim0.1$) are also reproduced
by the free approximation.

Third, we consider the Anderson model on a cubic lattice, whose Hamiltonian
is:

\begin{equation}
H^{3D}=\left(B^{1D}\otimes I\otimes I\right)+\left(I\otimes B^{1D}\otimes I\right)+\left(I\otimes I\otimes B^{1D}\right)+A
\end{equation}
Figure~\ref{fig:alldata}(e) shows the approximate and exact density
of states calculated from 1000 samples of $1000\times1000$ matrices.
This represents a $10\times10\times10$ cubic lattice which is significantly
smaller in linear dimension than the previously considered lattices.
We therefore observed oscillatory features in the density of states
arising from finite-size effects. Despite this, the free approximant
is still able to reproduce the exact density of states quantitatively.
In fact, if the histogram in Figure~\ref{fig:alldata}(e) is recomputed
with finer histogram bins to emphasize the finite-size induced oscillations,
we still observe that the free approximant reproduces these features.

\subsection{Off-diagonal disorder\label{sub:odd}}

$ $Up to this point, all of the models we have considered have only
site disorder, with no off-diagonal disorder. Free probability has
thus far provided a qualitatively correct approximation for all these
lattices. To test the robustness of this approximation, we now investigate
systems with random interactions. The simplest such system is the
one-dimensional chain, with a Hamiltonian of the form:

\begin{equation}
H_{ij}=g_{i}\delta_{ij}+h_{i}\left(\delta_{i,j+1}+\delta_{i,j-1}\right).\label{eq:1dodd}
\end{equation}
Unlike in the previous systems, the interactions are no longer constant,
but are instead new random variables $h_{i}$. We choose them to be
Gaussians of mean $J$ and variance $\left(\sigma^{*}\right)^{2}$.
There are now two order parameters to consider: $\sigma^{*}/J$, the
relative disorder in the interaction strengths, and $\sigma^{*}/\sigma$,
the strength of off-diagonal disorder relative to site disorder. As
in the prior one-dimensional case, we average over 1000 realizations
of $1000\times1000$ matrices.

We now observe that the quality of the free approximation is no longer
uniform across all values of the order parameters. Instead, it varies
with $\sigma^{*}/\sigma$, but not $\sigma^{*}/J$. In Figure~\ref{fig:alldata}(f),
we demonstrate the results of varying $\sigma^{*}/\sigma$ with $\sigma/J=1$.
In the limits $\sigma^{*}/\sigma\gg1$ and $\sigma^{*}/\sigma\ll1$,
the free approximation matches the exact result well; however, there
is a small but noticeable discrepancy between the exact and approximate
density of states for moderate relative off-diagonal disorder, though
the quality of the approximation is mostly unaffected by the centering
of the off-diagonal disorder. In the next section, we will investigate
the nontrivial behavior of the approximation with the $\sigma^{*}/\sigma$
order parameter.

\section{Error analysis \label{sub:error}}

In our numerical experiments, we have found that the accuracy of the
free approximation remains excellent for systems with only site disorder,
regardless of the underlying lattice topology or the number of interactions
that each site has. Details such as finite-size oscillations and Van
Hove singularities are also captured when present. However, when off-diagonal
disorder is present, the quality of the approximation does vary qualitatively
with the ratio of off-diagonal disorder to site disorder $\sigma^{*}/\sigma$
as illustrated in Section~\ref{sub:odd}, and the error is greatest
when $\sigma^{*}\approx\sigma$. To understand the reliability of
the free approximant (\ref{eq:free-approximant}) in all these situations,
we apply an asymptotic moment expansion to calculate the leading order
error terms for the various systems. In general, a probability density
$\rho$ can be expanded with respect to another probability density
$\tilde{\rho}$ in an asymptotic moment expansion known as the Edgeworth
series~\cite{Chen,Stuart1994}:
\begin{equation}
\rho\left(x\right)=\exp\left(\sum_{m=1}^{\infty}\frac{\kappa^{\left(m\right)}-\tilde{\kappa}^{\left(m\right)}}{m!}\left(-\frac{d}{dx}\right)^{m}\right)\tilde{\rho}\left(x\right)
\end{equation}
where $\kappa^{\left(m\right)}$ is the $m$th cumulant of $\rho$
and $\tilde{\kappa}^{\left(m\right)}$ is the $m$th cumulant of $\tilde{\rho}$.
When all the cumulants exist and are finite, this is an exact relation
that allows for the distribution $\tilde{\rho}$ to be systematically
corrected to become $\rho$ by substituting in the correct cumulants.
If the first $\left(n-1\right)$ cumulants of $\rho$ and $\tilde{\rho}$
match, but not the $n$th, then we can calculate the leading-order
asymptotic correction to $\tilde{\rho}$ as:\begin{subequations}

\begin{align}
\rho(x) & =\exp\left(\frac{\kappa^{\left(n\right)}-\tilde{\kappa}^{\left(n\right)}}{n!}\left(-\frac{d}{dx}\right)^{n}+\dots\right)\tilde{\rho}(x)\\
 & =\left(1+\frac{\kappa^{\left(n\right)}-\tilde{\kappa}^{\left(n\right)}}{n!}\left(-\frac{d}{dx}\right)^{n}+\dots\right)\tilde{\rho}(x)\\
 & =\tilde{\rho}\left(x\right)+\frac{(-1)^{n}}{n!}\left(\kappa^{\left(n\right)}-\tilde{\kappa}^{\left(n\right)}\right)\frac{d^{n}\tilde{\rho}}{dx^{n}}(x)+\mathcal{O}\left(\frac{d^{n+1}\tilde{\rho}}{dx^{n+1}}\right)\\
 & =\tilde{\rho}\left(x\right)+\frac{(-1)^{n}}{n!}\left(\mu^{\left(n\right)}-\tilde{\mu}^{\left(n\right)}\right)\frac{d^{n}\tilde{\rho}}{dx^{n}}(x)+\mathcal{O}\left(\frac{d^{n+1}\tilde{\rho}}{dx^{n+1}}\right)\label{eq:edgeworth}
\end{align}
\end{subequations}where on the second line we expanded the exponential
$e^{X}=1+X+\dots$, and on the fourth line we used the well-known
relationship between cumulants $\kappa$ and moments $\mu$ and the
fact that the first $n-1$ moments of $\rho$ and $\tilde{\rho}$
were identical by assumption.

We can now use this expansion to calculate the leading-order difference
between the exact density of states $\rho_{H}=\rho_{A+B}$, and its
free approximant $\rho_{H^{\prime}}=\rho_{A\boxplus B}$ by setting
$\tilde{\rho}=\rho_{H^{\prime}}$ and $\rho=\rho_{H}$ in (\ref{eq:edgeworth}).
The only additional data required are the moments $\mu_{H}^{\left(n\right)}=\left\langle H^{n}\right\rangle $
and $\mu_{H^{\prime}}^{\left(n\right)}=\left\langle \left(H^{\prime}\right)^{n}\right\rangle $,
which can be computed from the sampled data or recursively from the
joint moments of $A$ and $B$ as detailed elsewhere~\cite{Chen}.
This then gives us a way to detect discrepancies, which is to calculate
successively higher moments of $H$ and $H^{\prime}$ to determine
whether the difference in moments is statistically significant, and
then for the smallest order moment that differs, calculate the correction
using (\ref{eq:edgeworth}).

The error analysis also yields detailed information about the source
of error in the free approximation. The $n$th moment of $H$ is given
by 
\begin{equation}
\mu_{H}^{(n)}=\left\langle H^{n}\right\rangle =\left\langle \left(A+B\right)^{n}\right\rangle =\sum_{\begin{array}{c}
{\scriptstyle m_{1},n_{1},\dots,m_{k},n_{k}}\\
{\scriptstyle \sum_{j=1}^{k}m_{j}+n_{j}=n}
\end{array}}\left\langle A^{m_{1}}B^{n_{1}}\cdots A^{m_{k}}B^{n_{k}}\right\rangle ,
\end{equation}
where the last equality arises from expanding $\left(A+B\right)^{n}$
in a noncommutative binomial series. If $A$ and $B$ are freely independent,
then each of these terms must satisfy recurrence relations that can
be derived from the definition (\ref{eq:cjm})~\cite{Chen}. Exhaustively
enumerating and examining each of the terms in the final sum to see
if they satisfy (\ref{eq:cjm}) thus provides detailed information
about the accuracy of the free approximation.

We now apply this general error analysis for the specific systems
we have studied. It turns out that the results for systems with and
without off-diagonal disorder exhibit different errors, and so are
presented separately below.

\subsection{Systems with constant interactions}

We have previously shown that for the one-dimensional chain with nearest-neighbor
interactions, the free approximant is exact in the first seven moments,
and that the only term in the eighth moment that differs between the
free approximant and the exact $H$ is $\left\langle \left(AB\right)^{4}\right\rangle $~\cite{Chen2012}.
The value of this joint moment can be understood in terms of discretized
hopping paths on the lattice~\cite{Wigner1967}. Writing out the
term $\left\langle \left(AB\right)^{4}\right\rangle $ explicitly
in terms of matrix elements and with Einstein's implicit summation
convention gives:\begin{widetext}\begin{subequations}
\begin{align}
\left\langle \left(AB\right)^{4}\right\rangle  & =\frac{1}{N}\mathbb{E}\left(A_{i_{1}i_{2}}B_{i_{2}i_{3}}A_{i_{3}i_{4}}B_{i_{4}i_{5}}A_{i_{5}i_{6}}B_{i_{6}i_{7}}A_{i_{7}i_{8}}B_{i_{8}i_{1}}\right)\\
 & =\frac{1}{N}\mathbb{E}\left(\left(g_{i_{1}}\delta_{i_{1}i_{2}}\right)\left(JM_{i_{2}i_{3}}\right)\left(g_{i_{3}}\delta_{i_{3}i_{4}}\right)\left(JM_{i_{4}i_{5}}\right)\left(g_{i_{5}}\delta_{i_{5}i_{6}}\right)\left(JM_{i_{6}i_{7}}\right)\left(g_{i_{7}}\delta_{i_{7}i_{8}}\right)\left(JM_{i_{8}i_{1}}\right)\right)\\
 & =\frac{1}{N}\mathbb{E}\left(g_{i_{1}}g_{i_{2}}g_{i_{3}}g_{i_{4}}J^{4}M_{i_{1}i_{2}}M_{i_{2}i_{3}}M_{i_{3}i_{4}}M_{i_{4}i_{1}}\right).
\end{align}
\end{subequations}\end{widetext} From this calculation, we can see
that each multiplication by $A$ weights each path by the site energy
of a given site, $g_{i}$, and each multiplication by $B$ weights
the path by $J$ and causes the path to hop to a coupled site. The
sum therefore reduces to a weighted sum over returning paths on the
lattice that must traverse exactly three intermediate sites. The only
paths on the lattice with nearest-neighbors that satisfy these constraints
are shown in Figure~3(a), namely $\left(i_{1},i_{2},i_{3},i_{4}\right)=\left(k,k+1,k,k+1\right)$,
$\left(k,k+1,k+2,k+1\right)$, $\left(k,k-1,k,k-1\right)$, and $\left(k,k-1,k-2,k-1\right)$
for some starting site $k$. The first path contributes weight $\mathbb{E}\left(g_{k}^{2}g_{k+1}^{2}\right)J^{4}=\mathbb{E}\left(g_{k}^{2}\right)\mathbb{E}\left(g_{k+1}^{2}\right)J^{4}=\sigma^{4}J^{4}$
while the second term has weight $\mathbb{E}\left(g_{k}g_{k+1}^{2}g_{k+2}\right)J^{4}=\mathbb{E}\left(g_{k}\right)\mathbb{E}\left(g_{k+1}^{2}\right)\mathbb{E}\left(g_{k+2}\right)J^{4}=0$.
Similarly, the third and fourth paths also have weight $\sigma^{4}J^{4}$
and 0 respectively. Finally averaging over all possible starting sites,
we arrive at the final result that $\left\langle \left(AB\right)^{4}\right\rangle =2\sigma^{4}J^{4}$
with periodic boundary conditions and $\left\langle \left(AB\right)^{4}\right\rangle =2\left(1-1/N\right)\sigma^{4}J^{4}$
with vanishing boundary conditions. We therefore see when $N$ is
sufficiently large, the boundary conditions contribute a term of $\mathcal{O}\left(1/N\right)$
which can be discarded, thus showing the universality of this result
regardless of the boundary conditions.

\begin{figure}[h]
\caption{\label{fig:hopping}(a) Diagrammatic representation of the four paths
that contribute to the leading order error for the case of a two-dimensional
square lattice with constant interactions and nearest neighbors. Dots
contribute a factor of $g_{i}$ for site $i$. Solid arrows represent
a factor of $J$. Each path contributes $J^{4}\left\langle g_{a}^{2}\right\rangle \left\langle g_{b}^{2}\right\rangle =\sigma^{4}J^{4}$
to the error. (b) Build up of the diagrammatic representation the
leading order error in the case of a 1D chain with off-diagonal disorder.
The two dashed arrows contribute a factor of $\mu_{4}-\mu_{2}^{2}$.
Because of the disorder in the interactions, multiplication by $\overline{B^{2}}$
allows loops back to the same site. The first of these loops, $\left\langle \overline{A}\overline{B^{2}}\right\rangle $,
has zero expectation value because it contains an independent random
variable of mean zero as a factor. Once two loops are present, the
expectation value instead contains this random variable squared, which
has nonzero expectation value. }

\includegraphics[width=2.5in]{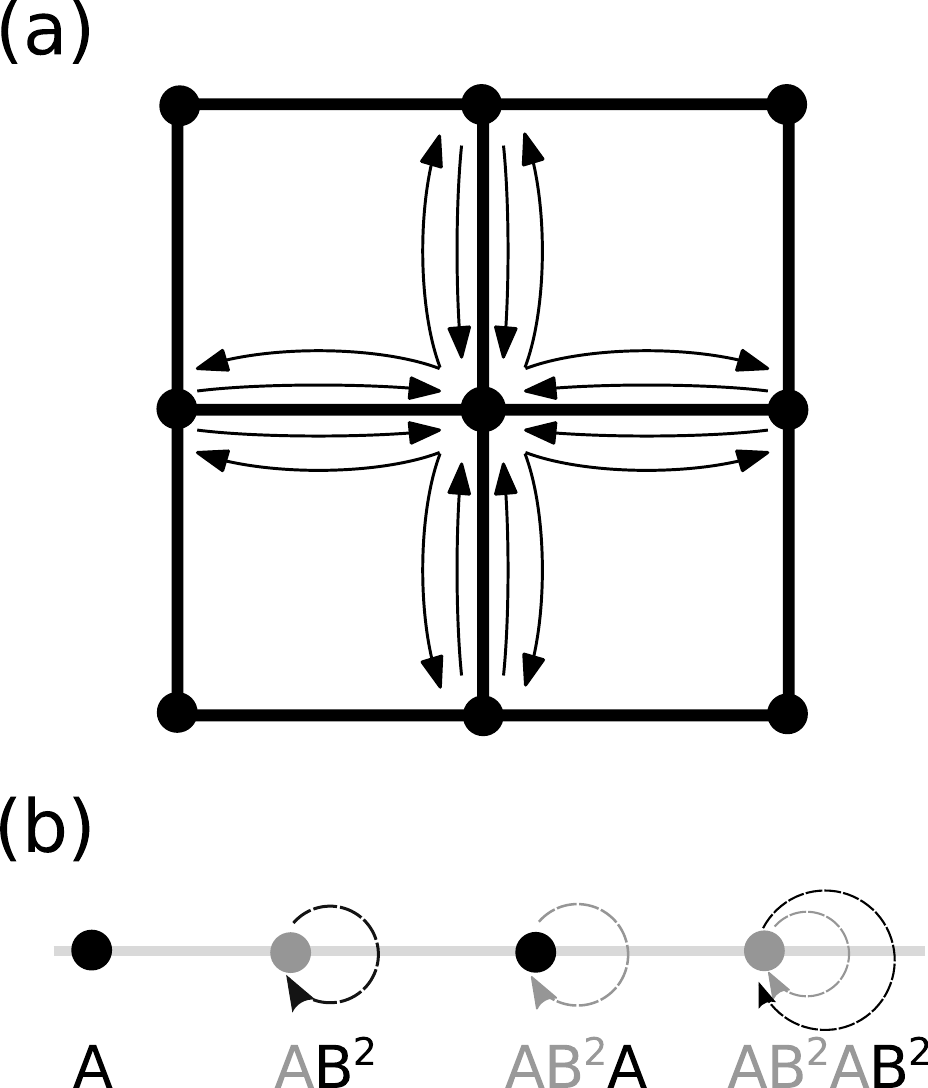}
\end{figure}

Applying the preceding error analysis, we observe that the result
from the one-dimensional chain generalizes all the other systems with
constant interactions that we have studied; the only difference being
that the coefficient 2 is simply replaced by $n$, the number of sites
accessible in a single hop from a given lattice site. In order to
keep the effective interaction felt by a site constant as we scale
$n$, we can choose $J$ to scale as $\frac{1}{\sqrt{n}}$. In this
case, the free approximation converges to the exact result as $\frac{1}{n}$.

We can generalize the argument presented above to explain why $ $$\left\langle \left(\overline{A}\overline{B}\right)^{4}\right\rangle $
is the first nonzero joint centered moment, and thus why the approximation
does not break down before the eighth moment. Consider centered joint
moments of the form:

\begin{equation}
\left\langle \overline{A^{a_{1}}}\overline{B^{b_{1}}}\overline{A^{a_{2}}}\overline{B^{b_{2}}}\ldots\overline{A^{a_{n}}}\overline{B^{b_{n}}}\right\rangle 
\end{equation}
for positive integers $\{a_{i},b_{i}\}$ such that $ $$\sum_{i}(a_{i}+b_{i})\leq8$.
Since $A$ is diagonal with iid elements, all powers of $A^{n}$ are
also diagonal with iid elements, and so $\overline{A^{n}}=0$. Centered
higher powers of $B$, $\overline{B^{n}}$, couple each site to other
sites with interaction strengths $J^{n}$, but after centering, the
diagonal elements of $\overline{B^{n}}$ are zero and multiplication
by $\overline{B^{n}}$ still represents a hop from one site to a different
coupled site. Therefore, the lowest order nonzero joint centered moment
requires at least four hops, so $n\ge4$ is the smallest possible
nonzero term, but but the only term of this form of eighth order or
lower is the one with $a_{i}=b_{i}=1$, i.e.\ the term $\left\langle \left(AB\right)^{4}\right\rangle $.

\subsection{Random interactions\label{sub:error-odd}}

When the off-diagonal interactions are allowed to fluctuate, the free
approximation breaks down in the sixth moment, where the joint centered
moment $\left\langle \left(\overline{A}\overline{B^{2}}\right)^{2}\right\rangle $
fails to vanish. We can understand this using a generalization of
the hopping explanation from before. In this case, $\overline{B^{2}}$
contains nonzero diagonal elements, which corresponds to a nonzero
weight for paths that stay at the same site. Thus, $\left(\overline{A}\overline{B^{2}}\right)^{2}$
contains a path of nonzero weight that starts at a site and loops
back to that site twice (shown in Figure~3(b)). The overall difference
in the moment of the exact distribution from that in the free distribution
is $2\sigma^{2}\left(\mu_{4}-\mu_{2}^{2}\right)$, where $\mu_{4}$
and $\mu_{2}$ are the fourth and second moments of the off-diagonal
disorder. As above, the $\sigma^{2}$ component of this difference
can be understood as the contribution of the two $A$s in the joint
centered moment. The other factor, $2\left(\mu_{4}-\mu_{2}^{2}\right)$,
is the weight of the path of two consecutive self-loops. The sixth
moment is the first to break down because, as before, we must hop
to each node on our path twice in order to avoid multiplying by the
expectation value of mean zero, and $\left(AB^{2}\right)^{2}$ is
the lowest order term that allows such a path.

We summarize the the leading order corrections and errors in Table~\ref{tab:errors}.
At this point, we introduce the quantity $\tilde{J}=\sqrt{2n}J$,
which is an aggregate measure of the interactions of any site with
all its $2n$ neighbors. As can be seen, the discrepancy occurs to
eighth order for all the studied systems with constant interactions,
with a numerical prefactor indicative of the coordination number of
the lattice, and the factor of 1/8! strongly suppresses the contribution
of the error terms. Furthermore, for any given value of the total
interaction $\tilde{J}$, the error decreases quickly with coordination
number $2n$, suggesting that the free probability approximation is
exact in the mean field limit of $2n\rightarrow\infty$ neighbors.
This is consistent with previous studies of the Anderson model employing
the coherent potential approximation.\cite{Thouless1974,Neu1995a,Neu1995b}
In contrast, the system with off-diagonal disorder has a discrepancy
in the sixth moment, which has a larger coefficient in the Edgeworth
expansion (\ref{eq:edgeworth}). This explains the correspondingly
poorer performance of our free approximation for systems with off-diagonal
disorder. Furthermore, the preceding analysis shows that only the
first and second moments of the diagonal disorder $\sigma$ contribute
to the correction coefficient, thus showing that this behavior is
universal for disorder with finite mean and standard deviation.

\begin{table}[h]
\caption{\label{tab:errors}Coefficients of the leading-order error in the
free probability approximation in the Edgeworth expansion (\ref{eq:edgeworth}).}

\centering{}%
\begin{tabular}{|r|c|c|c|}
\hline 
 & Order & Term & Coefficient\tabularnewline
\hline 
\hline 
1D & 8 & $\left(AB\right)^{4}$ & $\tilde{J}^{4}\sigma^{4}/\left(2\cdot8!\right)$\tabularnewline
\hline 
2D square & 8 & $\left(AB\right)^{4}$ & $\tilde{J}^{4}\sigma^{4}/\left(4\cdot8!\right)$$ $\tabularnewline
\hline 
2D honeycomb & 8 & $\left(AB\right)^{4}$ & $\tilde{J}^{4}\sigma^{4}/\left(3\cdot8!\right)$\tabularnewline
\hline 
3D cube & 8 & $\left(AB\right)^{4}$ & $\tilde{J}^{4}\sigma^{4}/\left(6\cdot8!\right)$\tabularnewline
\hline 
1D with $n$ nearest-neighbors & 8 & $\left(AB\right)^{4}$ & $\tilde{J}^{4}\sigma^{4}/\left(2n\cdot8!\right)$\tabularnewline
\hline 
1D with off-diagonal disorder & 6 & $\left(AB^{2}\right)^{2}$ & $\sigma^{2}\left(\mu_{4}-\mu_{2}^{2}\right)/6!$\tabularnewline
\hline 
\end{tabular}
\end{table}

\section{Conclusion}

Free probability provides accurate approximations to the density of
states of a disordered system, which can be constructed by partitioning
the Hamiltonian into two easily-diagonalizable ensembles and then
free convolving their densities of states. Previous work~\cite{Chen2012}
showed that this approximation worked well for the one-dimensional
Anderson model partitioned into its diagonal and off-diagonal components.
Our numerical and theoretical study described above demonstrates that
the same approximation scheme is widely applicable to a diverse range
of systems, encompassing more complex lattices and more interactions
beyond the nearest-neighbor. The quality of the approximation remains
unchanged regardless of the lattice as long as the interactions are
constant, with the free approximation being in error only in the eighth
moment of the density of states. When the interactions fluctuate,
the quality of the approximation worsens, but remains exact in the
first five moments of the density of states.

Our results strongly suggest that free probability has the potential
to produce high-quality approximations for the properties of disordered
systems. In particular, our theoretical analysis of the errors reveals
universal features of the quality of the approximation, with the error
being characterized entirely by the moments of the relevant fluctuations
and the local topology of the lattice. This gives us confidence that
approximations constructed using free probability will give us high-quality
results with rigorous error quantification. This also paves the way
for future investigations for constructing fast free convolutions
using numerical methods for $R$-transforms,\cite{Rao2012} which
would yield much faster methods for constructing free approximations.
Additionally, further studies will be required to approximate other
observables of interest such as conductivities and phase transition
points. These will require further theoretical investigation into
how free probability can help predict properties of eigenvectors,
which may involve generalizing some promising initial studies linking
the statistics of eigenvectors such as their inverse participation
ratios to eigenvalue statistics such as the spectral compressibility~\cite{Klesse1997,Bogomolny2011}.

\section*{Acknowledgements}

This work was funded by NSF SOLAR Grant No. 1035400. M.W. acknowledges
support from the NSF GRFP. We thank Alan Edelman, Eric Hontz, Jeremy
Moix, and Wanqin Xie for insightful discussions.

\bibliographystyle{apsrev4-1}
\bibliography{refs}

\end{document}